\documentclass[twocolumn,showpacs,preprintnumbers,amsmath,amssymb,prl,a4paper,superscriptaddress]{revtex4}
\usepackage{amssymb,amsmath,amsbsy}
\usepackage{mathrsfs}
\usepackage{graphicx}
\usepackage{epstopdf}
\usepackage{dcolumn}
\usepackage{bm}
\usepackage{verbatim}
\usepackage{subfigure}
\usepackage{color}
\usepackage{braket}

\newcommand{\gmuk}{g_{\mu,k}}
\newcommand{\gok}{g_{\text{o},k}}

\newcommand{\dmuk}{\delta_{\mu,k}}
\newcommand{\dok}{\delta_{\text{o},k}}

\newcommand{\Ok}{\Omega_k}

\newcommand{\sk}[1]{\sigma_{#1,k}}

\newcommand{\omegamu}{\omega_{b}}
\newcommand{\omegao}{\omega_a}
\newcommand{\omegadri}{\omega_\Omega}

\definecolor{green2}{rgb}{0,0.5,0}
\begin{document}
\title{A magneto-optic modulator with unit quantum efficiency}

\author{Lewis A. Williamson}
\author{Yu-Hui Chen}
\author{Jevon J. Longdell}
\email{jevon.longdell@otago.ac.nz}
\affiliation{Jack Dodd Centre for Photonics and Ultra-Cold Atoms, Department of Physics, University of Otago, Dunedin, New Zealand.}

\date{\today}
\begin{abstract}
We propose a device for the reversible and quiet conversion of microwave photons to optical sideband photons, that can reach 100\% quantum efficiency. The device is based on an erbium doped crystal placed in both an optical and microwave resonator. We show that efficient conversion can be achieved so long as the product of the optical and microwave cooperativity factors can be made large. We argue that achieving this regime is feasible with current technology and we discuss a possible implementation.
\end{abstract}
\maketitle

In recent years there has been spectacular progress in the development
of devices based on superconducting qubits for quantum information processing
\cite{clar08,nakamura,nakamura2,berk03,yama03,dica09,dica10,neel10}.
However two problems hinder the application of superconducting qubits,
namely the inability to send quantum states over long distances, and
the lack of a long term memory.  These two problems have spawned the
new field of `hybrid quantum systems', in which the coupling of
superconducting qubits to a wide range of other physical systems, such
as spin systems \cite{zhu11} and nano-mechanical systems
\cite{lahaye2009}, is being investigated. The two problems could also be
solved by the reversible inter-conversion between qubits encoded in
microwave photons (which couple naturally to superconducting qubits)
and optical photons.  The quantum memories that are available for
light \cite{hedg10,hoss09,stopped,usma10,timo13} could then be used,
as could optical fibers for the long distance transmission of quantum
states. 

There have been theoretical proposals \cite{prl1,prl2,prl3} and impressive experimental \cite{andrews2014} demonstrations of reversible and efficient but noisy conversion of microwave photons to optical sideband photons using a microwave and an optical resonator both coupled to the same  nano-mechanical oscillator. Here we propose inter-conversion between microwave photons and light by operating close to the narrow resonances in rare earth doped solids and using the resulting large nonlinearities. The advantage using rare earth dopants for the nonlineariety is that we only require temperatures cold enough to freeze out microwave frequency excitations rather than the very low temperatures required to freeze out the mechanical resonances. The narrow mechanical resonance also restricts the conversion bandwidth for nano-mechanical approaches. 
Frequency conversion from microwave to optical frequencies using doubly and triply resonant electro-optic modulators using conventional nonlinear crystals has been investigated for high efficiency conversion  \cite{cqeo1,cqeo2,ilch02} but are still a long way off unit quantum efficiency.

There have been a number of investigations of cavity QED using rare earth ion dopants, using either microwave \cite{stau12,prob13,tkalcec2014,probst2014} or optical \cite{ichi06,mcau11a} transitions. Here we propose building on this work by placing an erbium doped crystal in both a microwave and optical resonator. One nice feature of using erbium is that the photons produced will have a wavelength near 1540$\,$nm, where quantum states can been sent many kilometers over optical fiber.

\begin{figure}
  \centering
  \includegraphics[width=0.45\textwidth]{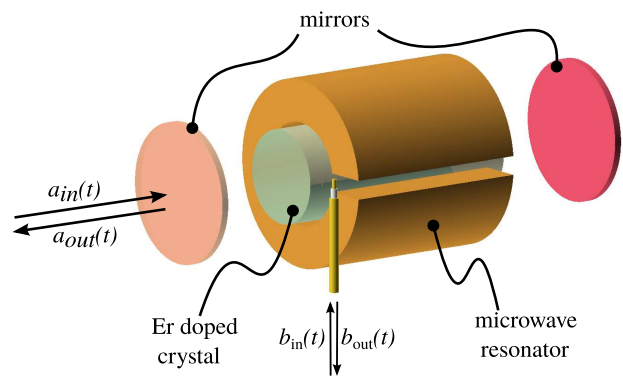}
  \caption{An outline of a device of the type we consider. An erbium doped crystal acts as an ensemble of $\Lambda$ systems and is coupled to a microwave resonator, an optical resonator and coherent driving field. To enable phase matching the coherent driving field will be applied using another mode of the optical resonator separated by one free spectral range.\label{fig:schem}}
\end{figure}

\begin{figure}
  \centering
  \includegraphics[width=0.2\textwidth]{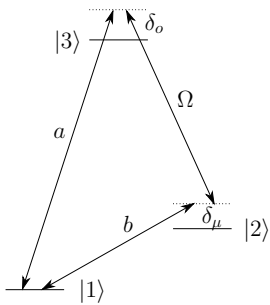}
  \caption{Energy level diagram of an erbium atom showing driven transitions. The microwave resonator is coupled to a spin transition. The optical resonator and coherent driving field drive the atoms from these two spin levels to a common excited state.}
  \label{fig:nrg}
\end{figure}

The setup we consider is shown diagrammatically in Fig.~\ref{fig:schem}. A collection of three level atoms interact with an optical cavity mode (frequency $\omega_a$), a microwave cavity mode (frequency $\omega_b$) and a coherent driving field (frequency $\omega_\Omega$), as shown in Fig.~\ref{fig:nrg}. The coupling strengths for the $k$-th atom to the microwave and optical resonators are $\gmuk$ and $\gok$ respectively and the coherent driving field has Rabi frequency $\Omega_k$. 

This leads to the following Hamiltonian for our atoms-cavities system.
\begin{eqnarray}
  \label{eq:hamil}
  H_\textrm{sys}/\hbar &=& \sum_k \left(\dok\sk{33} +\dmuk\sk{22}\right)+\sum_k  \left(\Ok\sk{32} + \text{h.c.}\right)\nonumber\\
&\phantom{=}&+ \sum_k \left(\gmuk\sk{21}b+\text{h.c.}\right)
+ (\gok\sk{31}a+\text{h.c.})
\end{eqnarray}
where the sum is over the active atoms, $ \sigma_{jk} \equiv |j\rangle\langle k|$, h.c.\ denotes Hermitian conjugate,  $a$ is the lowering operator for the optical resonator field and $b$ is the lowering operator for the microwave resonator field. Because of inhomogeneous broadening in both the optical and the spin transitions the values of $\dok$ and $\dmuk$ will vary from atom to atom. We have neglected the decay of the states $|2\rangle$ and $|3\rangle$. 

All three fields are detuned from the respective resonances in the atoms but they are in three photon resonance, $ \omegamu + \omegadri = \omegao$.
  
Working off resonance is important because the microwave cavity mode will include spins that are not in the optical mode. Working off resonance means that our precious microwave photons don't excite these \emph{parasitic} atoms. It also greatly simplifies the dynamics of the device because we can adiabatically eliminate the atom dynamics.

For the case of cryogenic rare earth ion dopants both the optical and spin transitions are inhomogeneously broadened. In order to adiabatic eliminate the atom dynamics the cavity modes must be detuned from line center by more than the inhomogeneous linewidth, so for any given atom the detuning will be very much larger than its homogeneous linewidth and so the spontaneous emission rates will be small.

 Working with large detunings, where $|\dok |\gg |\gok| $,  $|\dmuk| \gg |\gmuk|$ and $|\dok\dmuk|\gg |\Ok|^2 $, enables the adiabatic elimination of the excited states of the atoms \cite{brion2007,cohen1992} and yields

\begin{eqnarray}\label{eq:Heff}
H_{\mathrm{eff}}&= \hbar & \sum_\text{k}\left(-\frac {\dmuk|\gok|^2} {\dok\dmuk-|\Ok|^2}a^\dagger a-\frac {\dok|\gmuk|^2} {\dok\dmuk-|\Ok|^2} b^\dagger b \right. \nonumber\\
  &\phantom{=}&+ \left. \frac {\Ok\gmuk\gok^*} {\dok\dmuk-|\Ok|^2} a^\dagger b+\frac {\Ok^*\gmuk^*\gok} {\dok\dmuk-|\Ok|^2} b^\dagger a   \right) \label{eq:2}
\end{eqnarray}

The Hamiltonian in Eq.~(\ref{eq:Heff}) has four terms. The first two are due to the off resonant atoms pulling the resonant frequencies of the two cavities. We will ignore these terms as they can easily be compensated for by tuning the two resonators. The third and fourth term are a linear coupling between the two modes with strength which we shall denote by $S$. Because of the conditions required for adiabatic elimination $S$ becomes
\begin{equation}
  S = \sum_k\frac {\Ok\gmuk\gok^*} {\dok\dmuk}
\end{equation}

Using the input-output formalism we can get the following relations between the microwave and optical cavity fields and their input modes \cite{coll84}.
\begin{eqnarray}
\begin{split}
  \dot{a} &= -iSb-\frac{\kappa_a}{2} a - \sqrt{\kappa_a}a_\textrm{in}(t) \\
  \dot{b} &= -iS^*a-\frac{\kappa_b}{2} b - \sqrt{\kappa_b}b_\textrm{in}(t)
\end{split}
\end{eqnarray}
Here  $\kappa_a$  and $\kappa_b$ are the decay rates for the two cavities. Fourier transforming this  and using the input output relations \cite{coll84} gives.


\begin{eqnarray} \label{eq:fre}
\begin{split}
\tilde a_{out}(\omega)&=\frac{4iS \sqrt{\kappa_a\kappa_b}}{4 |S|^2+(\kappa_a-2i\omega)(\kappa_b-2i\omega) } \tilde b_{in}(\omega) \\
&\phantom{=}+\frac{4|S|^2 -(\kappa_a+2i\omega)(\kappa_b-2i\omega)}{4 |S|^2+(\kappa_a-2i\omega)(\kappa_b-2i\omega) }\tilde a_{in}(\omega)\\
\tilde b_{out}(\omega)&=\frac{4iS^* \sqrt{\kappa_a\kappa_b}}{4 |S|^2+(\kappa_a-2i\omega)(\kappa_b-2i\omega) } \tilde a_{in}(\omega) \\
&\phantom{=}+\frac{4|S|^2 -(\kappa_a-2i\omega)(\kappa_b+2i\omega)}{4 |S|^2+(\kappa_a-2i\omega)(\kappa_b-2i\omega)} \tilde b_{in}(\omega)
\end{split}
\end{eqnarray}

\begin{figure}
  \centering
\includegraphics[width=0.46\textwidth]{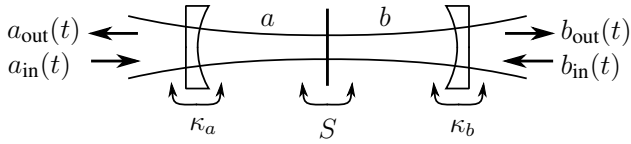}
  \caption{\label{fig:impedancematch} Two optical cavities that are
    coupled by sharing a partially transmissive end mirror. If the reflectivity of this end face mirror is tuned appropriately then, on resonance, the three mirrors will have 100\% transmission.}
\end{figure}

The first terms in the right-hand side of Eqns.~\eqref{eq:fre} give photon conversion between the microwave and optical fields and the second terms describe the signals reflected from the cavities. The number conversion efficiency is given by
\begin{equation}\label{eq:efficiency}
\begin{split}
\eta(\omega)=&\left|\frac{4iS \sqrt{\kappa_a\kappa_b}}{4 |S|^2+(\kappa_a-2i\omega)(\kappa_b-2i\omega) }\right|^2\\
\end{split}
\end{equation}
There is impedance matching when  $4|S|^2 = \kappa_a\kappa_b$ giving the desired result, on resonance $(\omega=0)$ the input microwave field is mapped completely onto the output optical field and vice-versa.  The bandwidth for the conversion is the geometric mean of the two cavity linewidths.
%
%
%
The situation is completely analogous to two optical cavities that share a partially transmissive end mirror. If the coupling between the two cavities is chosen appropriately the input to one of the cavities becomes the output of the other, see Fig.~\ref{fig:impedancematch}.

To get an intuitive understanding of the requirements for impedance matching we first assume that the $g$, $\Omega$, $\delta$, $\Delta$ parameters  are real and the same for each atom. By doing this we are ignoring for the moment the problems of phase matching and mode overlap by assuming that all the atoms are located at the maximum of the microwave and both the optical fields. With this assumption we can write our impedance matching condition $2|S|= \sqrt{\kappa_a\kappa_b}$ as 
\begin{equation}
\sqrt{
\frac{Ng^2_\mu}{\kappa_{b}\delta_\mu}
}
\times
\sqrt{
\frac{Ng^2_\text{o}}{{\kappa_{a}\delta_\text{o}}}
}
\times
\frac {2 \Omega}{ \sqrt{\delta_\mu \delta_\text{o}}}
= 1 \label{eq:crit}
\end{equation}

Obviously we can easily reduce the left hand side by turning down the classical drive field and therefore reducing $\Omega$. The challenge is to get the left hand side up to one.

In order that the cavity is detuned from the microwave resonance the $\delta_\mu$ needs to be bigger than the inhomogeneous linewidth ($\gamma_\mu$) of the spin transition  . This means that the first term in Eq.~\ref{eq:crit} is bounded above by the microwave cooperativity factor $\sqrt{{Ng_\mu^2}/({\kappa_b\gamma_\mu})}$. For an analogous reason the second term is bounded above by the optical cooperativity factor. The third term is bounded above by one due to the conditions for adiabatic elimination.
To achieve efficient upconversion it is therefore necessary to be in the strong coupling regime for either or both of the microwave and optical couplings, such that the product of the microwave and optical cooperativity factors is greater than one. This should be feasible in light of recent research in rare earth cavity QED. Using  microwave resonators and spin transitions  there are reports close to \cite{bushev2011,stau12} and achieving \cite{prob13,tkalcec2014,abe2011,probst2014,tabuchi14,zhang14} strong coupling. For optical cavity QED with rare earths  achieving the many atom strong coupling regime is straightforward, and people have strived for, but not yet achieved, strong coupling for a single dopant \cite{ichi06,mcau11a}. It is much easier to achieve strong coupling with many atoms because the penalty you pay, which is the square root of the ratio of inhomogeneous to homogeneous broadening, is much smaller than the benefit you get, which is the square root of the number of dopants. This is especially the case in systems where erbium replaces yttrium where the inhomogeneous broadening tends to be small.

We now relax the assumptions we made earlier to arrive at Eq.~\ref{eq:crit}. We allow the values of  $g$, $\Omega$, $\delta$, $\Delta$ again to vary from atom to atom. This is to account for imperfect phase matching and mode overlap between the optical and microwave modes as well as more accurately deal with inhomogeneous broadening.

To show that efficient conversion is feasible experimentally we will concentrate on three dimensional 3D copper microwave resonators containing the (nuclear spin free) even isotopes of erbium in yttrium orthosillicate (YSO). Er$^{3+}$ has an odd number of $f$ electrons, so for nuclear spin free isotopes both the lowest crystal field level of the ground state I$_{15/2}$ manifold and the lowest crystal field level of the electronically excited I$_{13/2}$ manifold are doubly degenerate. This degeneracy is broken by a magnetic field and 5~GHz ground state splitting can be achieved with a magnetic field of the order of 100\,mT. The effective spin Hamiltonian for the ground state and excited state are known \cite{sun2008}, allowing field orientations that lead to lambda transitions to be identified. 

An alternative approach to the one we consider here would be to use  $^{167}$Er, which has nuclear spin $7/2$, with superconducting microwave resonators. In YSO the hyperfine structure of  $^{167}$Er is split over $\sim5$\,GHz at zero magnetic field, which is attractive because superconducting resonators have high Q-factors at zero magnetic fields, but can suffer from additional losses in magnetic fields \cite{QinB}. The ground state hyperfine structure for $^{167}$Er:YSO has been determined \cite{guil06} and low frequency lambda transitions have been observed \cite{baldit2010} . However, the excited state hyperfine structure remains unknown and lambda systems using the full 5\,GHz splitting haven't been observed.

In order to  separate out the effects of inhomogeneous broadening from the spatial variations of the fields in $S$ we introduce two parameters. The first,  $\alpha$, only depends on the number density and spectroscopy of erbium,
\begin{equation}
\alpha\equiv\sqrt{\frac{\mu_0}{\hbar^2\epsilon_0 }}d_{31}\mu_{21}\rho\int_{\epsilon_\mu}^\infty\frac{D_\mu(\delta_\mu)}{\delta_\mu}\,d\delta_\mu\int_{\epsilon_\text{o}}^\infty\frac{D_\text{o}(\delta_\text{o})}{\delta_\text{o}}\,d\delta_\mu
\end{equation}
Here $d_{31}$ is the electric dipole moment for the $1\leftrightarrow 3$ transition, $\mu_{21}$ is the magnetic dipole moment for the $1\leftrightarrow 2$ transition, $\rho$ is the number density of the  Er ions within the crystal, and $D_\mu(\delta_\mu)$ and $D_\text{o}(\delta_\text{o})$ are inhomogeneous broadening distribution functions for the microwave and optical transitions respectively, which we shall assume to be Gaussian with standard deviations $\sigma_\mu$ and $\sigma_\text{o}$. The lower bounds on the integrals in $\alpha$ should be chosen far from the mean of the Gaussian distributions but need to be chosen $>0$ to avoid the problems at $\delta=0$ due to the breakdown of the adiabatic approximation.

The effects of imperfect phase matching and mode overlap can be described by a ``filling factor'' parameter  
\begin{equation}
F\equiv  \frac{1}{\sqrt{V_\mu V_\text{o}}}\left|\int_{V_\text{c}}\chi(\bm{r})\psi(\bm{r})\phi(\bm{r})\,d^3\bm{r}\right|
\end{equation}
Here $V_\text{c}$ is the crystal volume. The microwave and optical mode volumes are denoted by $V_\mu$ and $V_\text{o}$ respectively. The $\chi(\bm{r})$, $\psi(\bm{r})$ and $\phi(\bm{r})$ are the mode-functions 
for the microwave and two optical modes respectively. With these two parameters  we can write our impedance matching parameter $R$ as 
\begin{equation}
R\equiv \frac{2|S|}{\sqrt{\kappa_a \kappa_b}} =\Omega\alpha F\sqrt{Q_a Q_b}
\end{equation}
 where $\Omega$ is the peak Rabi frequency, and $Q_a$ and $Q_b$ are quality factors for the optical and microwave resonators respectively. 

\begin{figure}
  \centering
\includegraphics[width=0.5\textwidth]{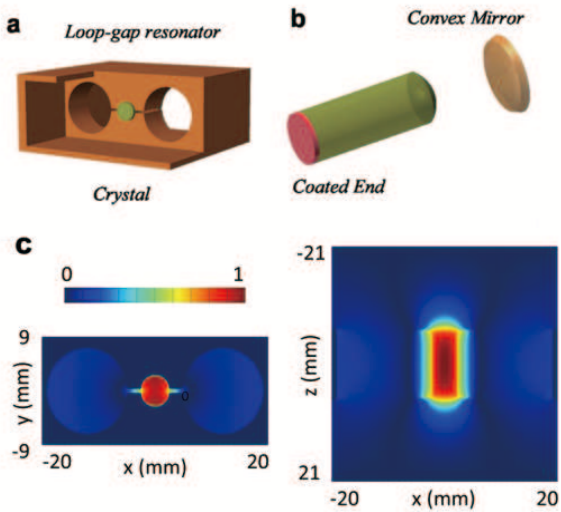}
  \caption{Diagrams showing proposed (a) microwave (to scale), (b) optical resonator, (not to scale) geometry and (c) microwave field distributions, on the left viewed from in front, on the right viewed from above. The 5\,mm-diameter crystal sits in the middle of the loop-gap resonator. The optical resonator would be hemilithic, with the anti reflection coated face of the crystal to improve the stability of the large mode diameter optical resonator\label{fig:resonator_design}}
\end{figure}

To consider what values for $R$ might be possible in practice we consider here a shielded loop-gap resonator \cite{lgr} and a Fabry-P\'erot resonator, as shown in Fig.~\ref{fig:resonator_design}. The magnetic field of the loop-gap resonator is concentrated in and reasonably uniform over the middle hole, see Fig.~\ref{fig:resonator_design}c. The optical modes are TEM$_{00}$ Gaussian modes with a waist diameter of 1\,mm, which is large but not unprecedented \cite{carstens2013}. We have made a number of microwave resonators similar to Fig.~\ref{fig:resonator_design} and have achieved $Q_b>2000$ and combined copper-dielectric resonators can have Q-factors as high as 70\,000 \cite{probst2014}.  The mode frequencies are $\omegamu=2\pi\times 5\,\text{GHz}$, $\omegao=2\pi\times 190\,\text{THz}$ and $\omegadri=\omegao-\omegamu$.
The value for $F$ for these resonators was calculated using FDTD (finite difference time domain) solutions for the microwave resonator and paraxial optics for the optical resonator, giving $F=0.0084$.

To calculate the $\alpha$ for our resonators we take $D_\mu$ to have a standard deviation of $1\,\text{MHz}$ and a mean of $3\sigma_\mu$ and $D_\text{o}$ to have a standard deviation of $500\,\text{MHz}$ and a mean of $3\sigma_\text{o}$ \cite{thiel2011}. We take $\epsilon_\mu=0.5\sigma_\mu$ and $\epsilon_\text{o}=0.5\sigma_\text{o}$.  For Er:YSO we have that $d_{31}=2.13\times 10^{-32}\,\text{Cm}$ \cite{mcau09} and for the $\ket{-15/2}\rightarrow \ket{15/2}$ spin transition we have that $\mu_{21}\approx 15\mu_\text{B}/2$, where $\mu_\text{B}$ is the Bohr magneton. The Zeeman $g$ tensor of Er:YSO is anisotropic and so maximizing the magnetic dipole moment requires correctly orientating the Er:YSO crystal \cite{prob13}.
We assume that the crystal is  a 0.001\% doped Er:YSO cylinder that fills the small hole of our loop-gap resonator. We then obtain that $\alpha=1.43\times 10^{-10}\,\text{s}$. We take $\Omega=10\,$MHz, which ensures that $\Omega^2<\delta_\mu\delta_\text{o}$ as required for the adiabatic approximation.

\begin{figure}[h]
\centering
\includegraphics[width=0.4\textwidth]{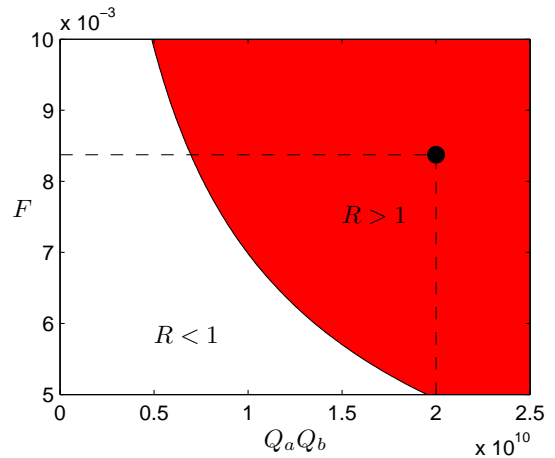}
\caption{\label{RPlot}Contour plot of $R$ versus the filling factor $F$ and the quality factor product $Q_a Q_b$. Complete photon conversion is achievable in the red region, where $R>1$. The horizontal dashed line shows the filling factor for the geometry shown in Fig.~\ref{fig:resonator_design} and the vertical dashed line corresponds to quality factors of $Q_a=10^7$ and $Q_b=2000$. The intersection of these lines gives a realistically achievable $R$ value of 1.7 (solid black disc) and therefore our resonator design is theoretically capable of achieving complete photon conversion.}
\end{figure}

A contour plot of $R$ versus $F$ and $Q_a Q_b$ provides a means to visualize the feasibility of achieving complete photon conversion and is shown in Fig.~\ref{RPlot}. Complete photon conversion is achievable in the red region where $R>1$.

Quality factors of $Q_a\gtrsim 10^8$ are obtainable for Fabry-Per\'{o}t resonators made out of YSO at 606\,nm \cite{goto2010}.  Taking $Q_a=10^7$ and $Q_b=2000$ gives $R=1.7$ and therefore our resonator design is theoretically capable of achieving complete photon conversion. It should be pointed out that there is room for improvement in our parameters. For example, isotopically pure erbium doped yttrium lithium fluoride has yielded 16~MHz  optical inhomogeneous linewidths \cite{erylf16MHz} rather than the 500~MHz used here. The microwave Q factors estimates used are also very conservative, with Q-factors of 70\,000 being demonstrated for some copper resonators \cite{probst2014}. 

Our theoretical analysis can be adapted to other cavity designs, such as a whispering gallery mode optical resonator in combination with a transmission line microwave resonator, similar to the most efficient electro-optic modulators demonstrated \cite{ilch02}.

In conclusion we propose using an erbium doped crystal in both an optical and microwave resonator to achieve complete photon conversion between microwave and optical fields. We present a theoretical analysis of a proposed design that should be within the reach of current technology. The analysis shows that our design is capable of achieving complete photon conversion.

The authors would like to acknowledge the Marsden Fund of the Royal Society of New Zealand for their support, and Rob Ballagh for his helpful comments on the manuscript. 

While preparing this manuscript we became aware of another proposal using erbium dopants for microwave upconversion that uses photon and spin echo techniques \cite{obri14}.


\end{document}